\documentclass[fleqn,usenatbib]{mnras}
\usepackage{newtxtext,newtxmath}
\usepackage[T1]{fontenc}
\usepackage{tablefootnote}
\DeclareRobustCommand{\VAN}[3]{#2}
\let\VANthebibliography\thebibliography
\def\thebibliography{\DeclareRobustCommand{\VAN}[3]{##3}\VANthebibliography}

\usepackage{graphicx}	% Including figure files
\usepackage{amsmath}
\usepackage{xspace} % Define commands that appear not to eat spaces
\renewcommand{\AA}{\normalfont\r{A}\xspace} % Following space is rendered, and in math mode show in rm font as units should be.

\usepackage{color,soul}
\definecolor{lightblue}{rgb}{.70,.95,1}
\sethlcolor{lightblue}

\newcommand{\teff}{\ensuremath{T_{\mathrm{eff}}}\xspace}

\newcommand{\kms}{\ensuremath{\rm{km}\,s^{-1}}\xspace}
\newcommand{\logg}{\ensuremath{\log g}\xspace}
\newcommand{\feh}{\rm{[Fe/H]}\xspace}
\newcommand{\cfe}{\rm{[C/Fe]}\xspace}
\newcommand{\nfe}{\rm{[N/Fe]}\xspace}

\newcommand{\ac}{\rm{A(C)}\xspace}

\newcommand{\alphafe}{\rm{[\ensuremath{\alpha}/Fe]}\xspace}
\newcommand{\Gaia}{\textit{Gaia}\xspace}
\newcommand{\CaHK}{\emph{CaHK}\xspace}
\newcommand{\Pristine}{\emph{Pristine}\xspace}
\newcommand{\FERRE}{{\tt FERRE}\xspace}
\newcommand{\ULySS}{{\tt ULySS}\xspace}

\def\msun{{\rm\,M_\odot}}
\def\eg{{e.g.,\ }}
\def\ie{{i.e.,\ }}

\title[GTC follow-up of Pristine/LAMOST]{The Pristine survey -- XX: GTC follow-up observations of extremely metal-poor stars identified from Pristine and LAMOST}

\author[Arentsen et al.]{Anke Arentsen,$^{1,2}$ \thanks{Email: \url{anke.arentsen@ast.cam.ac.uk}}
David S. Aguado,$^{3,4}$ 
Federico Sestito,$^{5}$
Jonay I. Gonz\'alez Hern\'andez,$^{6,7}$
\newauthor
Nicolas F. Martin,$^{2,8}$
Else Starkenburg,$^{9}$
Pascale Jablonka$^{10,11}$ and 
Zhen Yuan$^{2}$
\\ 
\\
$^{1}$ Institute of Astronomy, University of Cambridge, Madingley Road, Cambridge CB3 0HA, UK \\
$^{2}$ Universit\'e de Strasbourg, CNRS, Observatoire astronomique de Strasbourg, UMR 7550, F-67000 Strasbourg, France \\
$^{3}$ Dipartimento di Fisica e Astronomia, Universit\'a degli Studi di Firenze, Via G. Sansone 1, I-50019 Sesto Fiorentino, Italy \\
$^{4}$ INAF-Osservatorio Astrofisico di Arcetri, Largo E. Fermi 5, I-50125 Firenze, Italy. \\
$^{5}$ Department of Physics and Astronomy, University of Victoria, PO Box 3055, STN CSC, Victoria BC V8W 3P6, Canada \\
$^{6}$ Instituto de Astrof{\'\i}sica de Canarias, E-38205 La Laguna, Tenerife, Spain  \\
$^{7}$ Universidad de La Laguna, Dpto. Astrof{\'\i}sica, E-38206 La Laguna, Tenerife, Spain \\
$^{8}$ Max-Planck-Institut f\"ur Astronomie, K\"onigstuhl 17, D-69117 Heidelberg, Germany\\
$^{9}$ Kapteyn Astronomical Institute, University of Groningen, Postbus 800, 9700 AV, Groningen, the Netherlands\\
$^{10}$ Laboratoire d’astrophysique, École Polytechnique Fédérale de Lausanne (EPFL), Observatoire, CH-1290 Versoix, Switzerland \\
$^{11}$ GEPI, Observatoire de Paris, Université PSL, CNRS, Place Jules Janssen, F-92195 Meudon, France 
}

\date{Accepted 2023 January 05. Received 2023 January 05; in original form 2022 November 03 }

\pubyear{2023}

\begin{document}
\label{firstpage}
\pagerange{\pageref{firstpage}--\pageref{lastpage}}
\maketitle

% Abstract of the paper (max 250 words)
\begin{abstract}
Ultra metal-poor stars ($\feh < -4.0$) are very rare, and finding them is a challenging task. Both narrow-band photometry and low-resolution spectroscopy have been useful tools for identifying candidates, and in this work we combine both approaches.
We cross-matched metallicity-sensitive photometry from the \Pristine survey with the low-resolution spectroscopic LAMOST database, and re-analysed all LAMOST spectra with $\feh_\Pristine<-2.5$. We find that $\sim$1/3rd of this sample (selected without $\feh_\Pristine$ quality cuts) also have spectroscopic $\feh < -2.5$. 
From this sample, containing many low signal-to-noise (S/N) spectra, we selected eleven stars potentially having $\feh < -4.0$ or $\feh < -3.0$ with very high carbon abundances, and we performed higher S/N medium-resolution spectroscopic follow-up with OSIRIS on the 10.4m Gran Telescopio Canarias (GTC). We confirm their extremely low metallicities, with a mean of $\feh = -3.4$ and the most metal-poor star having $\feh = -3.8$. 
Three of these are clearly carbon-enhanced metal-poor (CEMP) stars with $+1.65 < \cfe < +2.45$.
The two most carbon-rich stars are either among the most metal-poor CEMP-s stars or the most carbon-rich CEMP-no stars known, the third is likely a CEMP-no star.
We derived orbital properties for the OSIRIS sample and find that only one of our targets can be confidently associated with known substructures/accretion events, and that three out of four inner halo stars have prograde orbits. 
Large spectroscopic surveys may contain many hidden extremely and ultra metal-poor stars, and adding additional information from e.g. photometry as in this work can uncover them more efficiently and confidently. 
\end{abstract}


%%%%%%%%%%%%%%%%%%%%%%%%%%%%%%%%%%%%%%%%%%%%%%%%%%

%%%%%%%%%%%%%%%%% BODY OF PAPER %%%%%%%%%%%%%%%%%%

\section{Introduction}

The most metal-poor stars still present in the Milky Way today are valuable portals to the early Universe and the pristine environments these stars were born in. They are thought to have formed from material enriched by the first generation(s) of stars, and their chemical abundances can be used to constrain the properties of the stars that came before them. Additionally, the dynamical properties of the most metal-poor stars teach us about the early formation of the Milky Way. Much can be, and has been, learned from very/extremely/ultra metal-poor stars with $\feh < -2.0\,\mathrm{(VMP)}/-3.0\,\mathrm{(EMP)}/-4.0\,\mathrm{(UMP)}$ \citep[e.g.][]{beerschristlieb05, frebelnorris15}, although they are exceedingly rare. 

The metal-poor halo has been found to be a melting pot of many accreted structures. It is populated by the remnants of the larger mergers that the Galaxy experienced across its history, such as Gaia-Sausage/Enceladus \citep[GSE, \eg][]{Belokurov18,Helmi18}, Sequoia \citep[\eg][]{Barba19, myeong2019}, Thamnos \citep[\eg][]{Koppelman19}, and Sagittarius \citep[\eg][]{Ibata94}. The plethora of recently discovered stellar streams are indicative of part of the later accretion events from dwarf/ultra faint galaxies and globular clusters \citep[\eg][]{Ibata21,li22,Martin22a,Martin22}. Additionally, as much as half of the stars in the halo appears to be born \textit{in-situ}, likely consisting of both an $\alpha$-rich splashed disk component \citep[\eg][]{bonaca17, haywood18, dimatteo19, gallart19, belokurov20} and stars that formed in a hot and disordered pre-disk state \citep[\eg][]{Belokurov22, conroy22}. 

The common picture from various cosmological simulations suggests that the VMP stars that inhabit the spatial inner region of the Milky Way, \ie the bulge and the disk, are amongst the oldest stars \citep[\eg][]{Starkenburg17b, Elbadry18, Sestito21}.  
These stars are therefore great tracers of the early Galactic assembly. On the observational point of view, many VMP stars have been observed with such kinematics, focusing on the bulge \citep[\eg][]{Howes14, Howes15, Howes16, Arentsen20, Lucey22, Sestito23} and the disk \citep[\eg][]{Sestito19, Sestito20, DiMatteo20, Carter21, Cordoni21}. The chemical properties of these populations indicate that the building blocks of the inner Galaxy consisted of a variety of objects -- some stars appear to have formed in systems very similar to ultra faint dwarf galaxies, while others are consistent with being born in globular cluster-like systems \citep[\eg][and references therein]{Schiavon17,Sestito23}, and finally there may also be a significant contribution of \textit{in-situ} VMP stars in the inner Galaxy \citep{Belokurov22, rix22}. 

Many low-metallicity stars have been found to be carbon-enhanced metal-poor (CEMP) stars, with frequencies of the order of $30-50\%$ among stars with $\feh < -3.0$ \citep{beerschristlieb05, yong13, placco14}. There are two main types of CEMP stars. CEMP-s stars are thought to have become carbon-rich later in their life due to mass-transfer from a (former) asymptotic giant branch (AGB) star companion -- these are typically in binary systems \citep[e.g.][]{hansen16a}, are enhanced in s-process elements as well as carbon (a signature of AGB star nucleosynthesis), and are more frequent for $\feh > -3.0$. The CEMP-no stars are hypothesised to have been born from carbon-enhanced gas in the early Universe -- they do not have s-process over-abundances, are less frequently found to be in binary systems (e.g. \citealt{hansen16b}, although still more than expected, see \citealt{arentsen19}), and mostly occur at $\feh < -3.0$. The exact frequencies of CEMP-no and CEMP-s stars as function of metallicity is still under debate \citep{arentsen22}, and may also vary with Galactic environment (e.g. inner vs. outer halo, bulge, dwarf galaxies, globular clusters). 

To build large samples of extremely metal-poor stars, many dedicated searches have happened in the past 40 years. Several different techniques have been used to identify metal-poor stars, such as following up high-proper motion stars with ultraviolet excesses \citep{ryannorris91}, identifying objects with small Ca II H \& K lines in large objective-prism surveys \citep{beers85, christlieb08}, or using metallicity-sensitive (narrow-band) photometry \citep{schlaufmancasey14, starkenburg17, dacosta19, galarza22, placco22}. Very and extremely metal-poor stars have also been identified in greater numbers in large scale spectroscopic surveys such as the Sloan Digital Sky Survey \citep[SDSS,][]{york2000},  the Large sky Area Multi-Object fiber Spectroscopic Telescope \citep[LAMOST\footnote{\url{http://www.lamost.org/public/?locale=en}},][]{Deng12}, RAdial Velocity Experiment \citep[RAVE,][]{steinmetz06} and the GALactic Archaeology with HERMES spectroscopic survey \citep[GALAH,][]{buder21}, see e.g. \citet{lee13}, \citet{li18}, \citet{matijevic17} and \citet{hughes22}. These are often paired with dedicated follow-up efforts (\citealt{caff13I, alle15, bonifacio15, aguado16, placco18, li22,dacosta22}). 

In this work, we combine the strengths of metallicity-sensitive photometry and large spectroscopic surveys by cross-matching metal-poor candidates from the photometric \Pristine survey \citep{starkenburg17} with the large database of spectra from LAMOST, with the goal of identifying new extremely or even ultra metal-poor stars. The \Pristine survey uses metallicity-sensitive narrow-band \CaHK photometry to derive photometric metallicities of millions of stars towards the Galactic halo, which is very efficient even for extremely metal-poor stars \citep{youakim17, aguado19}. However, the selection still suffers from some more metal-rich contamination. In this work we alleviate this by adding an extra step, namely by cross-matching candidates with $\feh_\Pristine < -2.5$ with the LAMOST spectroscopic database, and doing a dedicated analysis of all these (often low signal-to-noise) spectra. We select exciting candidates from this analysis, and follow them up using the OSIRIS spectrograph at the 10.4m Gran Telescopio Canarias (GTC) \citep{cepa2000} to obtain higher S/N observations, from which we can derive high-quality metallicities to confirm their extremely metal-poor nature.

We describe our initial candidate selection from \Pristine and LAMOST in Section~\ref{sec:selection}, including some discussion about the success rates. The OSIRIS observations for 11 stars and the derivation of their radial velocities, stellar parameters, distances and orbits is described in Section~\ref{sec:osiris}. We present results for the OSIRIS sample in Section~\ref{sec:results}, discussing the presence of carbon-enhanced metal-poor (CEMP) stars, the orbital properties for the sample and a comparison with a new value-added LAMOST catalogue. We conclude in Section~\ref{sec:conclusions}.

\section{Selection of EMP candidates using Pristine and LAMOST}\label{sec:selection}

The LAMOST archive contains low-resolution spectra (R$\sim$1800) for millions of stars, but not all spectra have stellar parameters in the standard LAMOST catalogue tables. We discovered that many of the most metal-poor stars ($\feh < -2.5$) are missed by their standard pipeline \citep{wu14}, and also by the dedicated very metal-poor pipeline of \citet{li18}. This is particularly severe for hotter stars and stars with lower signal-to-noise (S/N). Other dedicated analyses might be able to deal better with these spectra, and identify promising extremely metal-poor stars. 

At the time our selection was made (February 2021), the latest LAMOST release was DR6. To avoid having to analyse the full data release, which contains almost 10 million spectra, we made a pre-selection of promising extremely metal-poor candidates using photometric metallicities from the \Pristine survey. We used the internal \Pristine data release containing all \CaHK observations until Semester 2020A, and adopted the \CaHK + SDSS photometric metallicities \citep{starkenburg17}. We queried the LAMOST archive for all stars in the \Pristine survey with photometric metallicities $\feh_\Pristine < -2.5$ (from using either $g-i$ or $g-r$) and $g_\mathrm{sdss} < 18$, and found $\sim 7500$ cross-matches for $\sim 6000$ unique targets. No other quality cuts were applied, which usually are included when we do dedicated target selection for \Pristine follow-up immediately from the photometry \citep{youakim17}, to be as inclusive as possible.

\subsection{Preliminary \ULySS analysis}

A first-pass analysis of these candidates was done with the \ULySS\footnote{ULySS is available from \url{http://ulyss.univ-lyon1.fr/}} code \citep{koleva09}. \ULySS is a full-spectrum fitting package that employs empirical spectral libraries to determine stellar parameters (\teff, \logg, \feh, radial velocities, spectral broadening), and can be applied to stars of a wide range of stellar parameters and metallicities. We employed this code because we were interested in the types of contamination in the \Pristine selection, which one cannot study with the dedicated metal-poor analysis described in the next sub-section. 

For the models, we adopted the empirical MILES library \citep{sanchez06, falcon11} and used the \ULySS MILES polynomial interpolator originally built by \citet{prugniel11} and updated for cool stars by \citet{sharma16}. The library has a resolving power of $R \sim 2200$, and the interpolator extends down to $\feh = -2.8$ (with the possibility to extrapolate, at ones own risk). The LAMOST spectra were fitted between 3750 and 5500 \AA using a multiplicative Legendre polynomial of degree 15 for the normalisation. This degree is large enough to absorb some of the large mismatches between models and observations for carbon-enhanced metal-poor stars in regions of carbon-related molecular bands, which is necessary since the ULySS models do not include \cfe as a free parameter, and large carbon features could mess up the normalisation. There is also an automatic masking routine in ULySS, which excludes outlier pixels iteratively and typically masks the wavelength regions of the largest carbon features in CEMP stars.

\begin{figure}
\centering
\includegraphics[width=1.0\hsize,trim={0.0cm 1.0cm 0.0cm 1.0cm}]{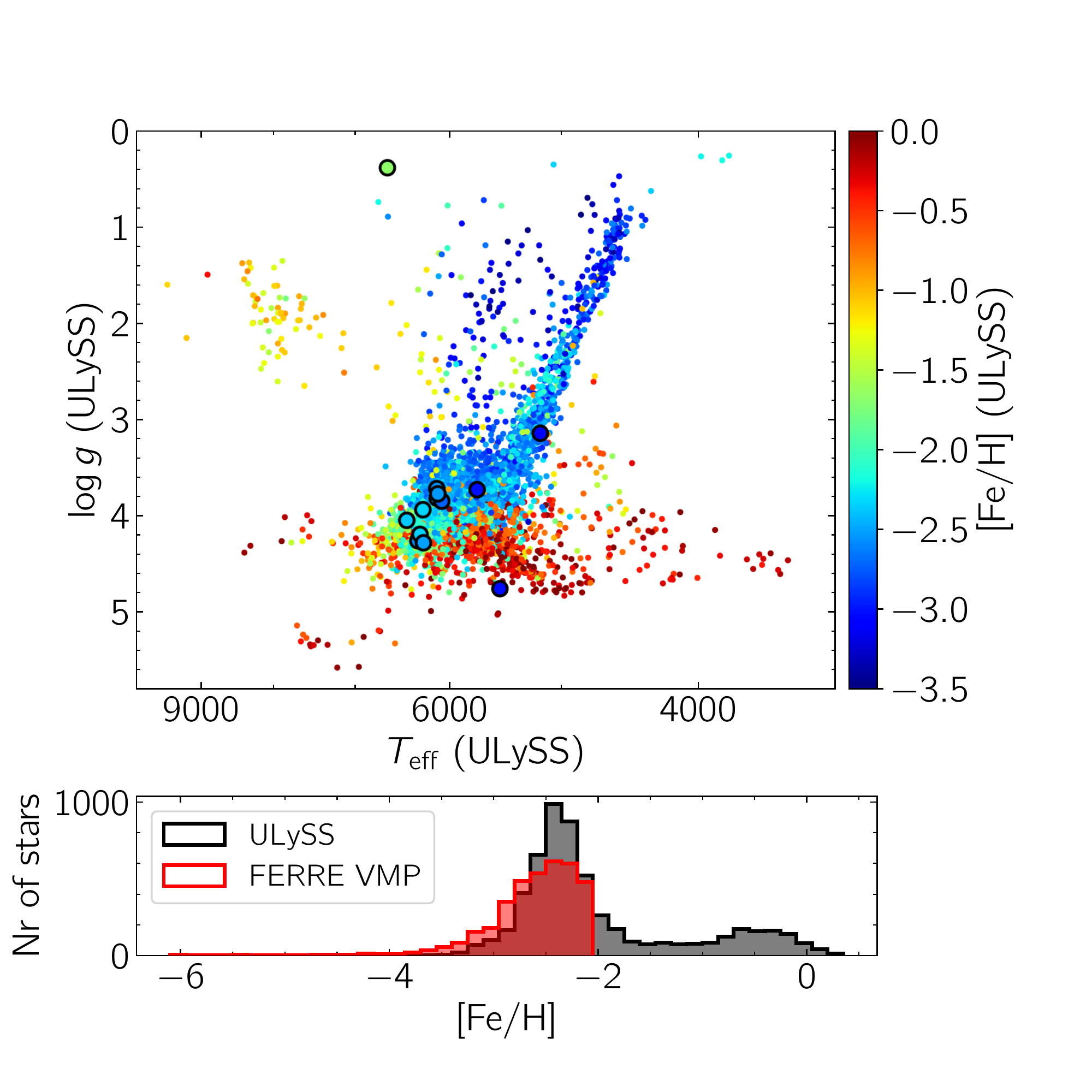}
\caption{Top: Kiel diagram for all exposures of the 4900 unique \Pristine-selected stars in LAMOST analysed with \ULySS, colour-coded by metallicity. No quality cuts were applied to the photometric metallicities in the selection. The results for the eleven stars that were followed up with OSIRIS (see Section~\ref{sec:osiris}) are highlighted with larger symbols (the two high and low \logg outliers are CEMP stars). Bottom: \ULySS metallicity histogram of the same sample in black, and \FERRE metallicity histogram for the VMP sub-sample in red. }
\label{fig:ulyss} 
\end{figure}

The resulting Kiel-diagram and metallicity histogram from our \ULySS analysis are shown in Figure~\ref{fig:ulyss}, for all exposures of the $\sim4900$ unique stars that remain after removing fits with signal-to-residual ratios $<8$, broadening $> 400$~\kms (which usually indicates a very bad fit), and duplicate LAMOST spectra for the same star. The metal-poor stars show a clear red giant branch (RGB) and main-sequence turn-off sequence, except for a small cloud of stars to the left of the RGB, which mostly consists of stars in the low S/N-tail of the sample without good fits.

Most of the stars in our selection are indeed very metal-poor (keeping only the fit with the highest signal-to-residual ratio per star): $71\%$ have $\feh_\ULySS < -2.0$ and $25\%$ have $\feh_\ULySS < -2.5$. The latter goes up to $34\%$ when using the \FERRE metallicities described later in this section, which perform better in this regime than the \ULySS metallicities, because the MILES library does not have many stars in this \feh range (especially for the turn-off region). There is a contamination of metal-rich stars with $\feh_\ULySS > -1.0$ of $16\%$. 

\ULySS is also the main software used by the LAMOST team for the parameters in their public data releases \citep{wu14}. They use the interpolator based on the ELODIE library \citep{prugnielsoubiran01, wu11}, which has a more limited coverage of the parameter space compared to MILES, and extends only down to $\feh = -2.5$. Of the stars that have $\feh_\ULySS < -2.0/-2.5$ in our analysis, only $30\%/17\%$ have stellar parameters in the public LAMOST DR7 catalogue. This is likely partly due to the ELODIE library being less good at low metallicities, and partly due to more stringent quality cuts being applied for stars to make it into the LAMOST data releases. 

\subsection{Success rates}

In our original selection we did not make any additional photometric quality cuts. The \Pristine team developed several quality cuts to remove metal-rich outliers and improve the success rates of the spectroscopic follow-up of extremely metal-poor candidates. The cuts applied for the main \Pristine follow-up campaign are discussed in Section~4.1 of \citet{youakim17}. We apply very similar cuts to the \Pristine+LAMOST sample to see how that changes the metallicity distribution, keeping only the stars that have: 

\begin{itemize}
    \item CASU flag = $-1$ or 1
    \item young stars flag = 0
    \item $(u_0 - g_0) > 0.6$
    \item $ 0.25 < (g_0 - i_0) < 1.5$ and $ 0.15 < (g_0 - r_0) < 1.2$
    \item $\feh_\Pristine < -2.5$ (from using either SDSS $g-i$ or $g-r$) and $\neq -99$ ($-99$ is assigned if the star falls outside of the parameter space for which the photometric metallicity assignment has a valid calibration)
    \item instead of the PanSTARRS variability catalogue as in \citet{youakim17}, we use the \Gaia photometric variability to remove variable stars as in \citet{fernandez21}
\end{itemize}

\begin{table*}
\begin{center}
\caption{List of 481 EMP candidates (533 spectra) with FERRE spectroscopic parameters used for target selection. No quality cuts have been applied. The first few lines of the table are shown here for guidance, the full table and figures showing the best fits are available as online supplementary material.}
\resizebox{\textwidth}{!}{
 \begin{tabular}{|c c c c c c c c c c|} 
 \hline
 LAMOST spectrum name & Gaia DR3 source\_id  &  ra & dec  & \teff & \logg & \feh & \cfe & S/N & $\log(\chi^2)$\\ 
 &   &  [deg] & [deg]  & [K] & cgs &  &  &  & \\ 
 \hline\hline

spec-56746-HD121251N314746M01\_sp02-087 & 4014278062082321152	& 181.215811 & 	30.331285  &	$6232 \pm 325$ & $5.0 \pm 1.4$	& $-3.7 \pm 15.1$	& $-0.4 \pm 90.4$	 & 6 & $-0.338$ \\
spec-57308-EG000023N024031M01\_sp03-076  &	2739551594198812160 & 358.800705	& 2.493245	& $6521	\pm 161$ & $4.7 \pm 0.4$ & $-3.2 \pm 0.8$ & $0.1 \pm 6.9$ & 24 & $-0.214$ \\
spec-57754-HD122624N271605M02\_sp03-108 & 4009964020835772288 & 185.239126	& 27.687022	& $5675 \pm 57$ &  $4.3 \pm 0.3$ & $-3.4 \pm 0.1$ & $0.3 \pm 0.3$ & 34 & $-0.123$ \\ 
 ... & ... & ... & ... & ... & ... & ... & ... & ... & ... \\
 [1.0ex]
 \hline
\end{tabular}}
\label{tab:allcandidates}
\end{center}
\end{table*}

The uncertainties on $\feh_\Pristine$ are not taken into account here, whereas they were in \citet{youakim17}. After applying the above cuts, the sample goes from 4900 stars to 4100 stars again keeping the highest signal-to-residual spectrum per star). Of these, $78\%$ have $\feh_\ULySS < -2.0$, and $28\%$ have $\feh_\ULySS < -2.5$ (the latter goes up to 38\% for the \FERRE metallicities), compared to the previous $71\%$ and $25\%$ (and $34\%$ for \FERRE), respectively. The metal-rich contamination goes down to $12\%$. Doing the same only for stars with signal-to-residual ratios $>20$ instead of our initial cut at $>8$, the results are very similar. We conclude that, for the \Pristine+LAMOST sample, the photometric quality cuts slightly improve the selection efficiency, but not by a lot. 

The success rate of previous \Pristine follow-up for $\feh_\Pristine < -2.5$ was found to be 56\% \citep{aguado19}. The lower fraction in this work (38\% when applying the photometric quality cuts and adopting the \FERRE metallicities) could be due to various reasons. For example, the dedicated \Pristine follow-up presented in \citet{youakim17} and \citet{aguado19} focused on the most metal-poor stars, and did not homogeneously observe \textit{all} stars with $\feh_\Pristine < -2.5$. Additionally, extra quality cuts were sometimes implemented for subsets of the follow-up presented in \citet{aguado19}, most notably the consistency of the photometric temperatures derived from SDSS between $(g-i)$ and $(g-r)$. If we select only stars with $\feh_\Pristine < -2.7$ and $|\teff\,_{(g-i)} - \teff\,_{(g-r)}| < 200$~K, the success rate in this work for $\feh_\FERRE < -2.5$ goes up to $50\%$, and the contamination of stars with $\feh_\ULySS > -1.0$ is reduced from $12\%$ to $2\%$. 

The numbers in this section are not meant to override the previously published \Pristine success rates by \citet{youakim17} and \citet{aguado19}. Our results confirm that the success rates are high, and highlight some of the subtleties in deriving such success rates. Overall we conclude that our methodology to find hidden very and extremely metal-poor stars in the large LAMOST database is extremely efficient.

\subsection{Dedicated VMP \FERRE analysis}\label{sec:fullferre}

A dedicated very metal-poor analysis was performed for the sub-sample of LAMOST spectra with $\feh_\ULySS < -2.0$, with the aim of deriving better metallicities and carbon abundances for the most metal-poor stars and identifying potential ultra metal-poor candidates. We followed a similar methodology as in \citet{agu17II, agu17I}, using the \FERRE\footnote{\FERRE is available from \url{http://github.com/callendeprieto/ferre}} code \citep{alle06}. The code interpolates between the nodes of a library of synthetic spectra and derives simultaneously the set of best stellar parameters (\teff, \logg, \feh, \cfe). For this preliminary analysis, we used the default Nelder-Mead search algorithm and linear interpolation. The dedicated very metal-poor synthetic models were computed with the {\tt ASSET} code \citep{koe08} and published in \citet{agu17I} with the following parameter coverage:

\begin{itemize}
\item $4750~\mathrm{K} < \teff < 7000~\mathrm{K}$, $\Delta \teff = 250~\mathrm{K}$
\item $1.0  < \logg < 5.0$, $\Delta \logg = 0.5$
\item $-6.0  < \feh < -2.0$, $\Delta \feh = 0.5$
\item $-1.0  < \cfe < +5.0$, $\Delta \cfe = 1.0$
\end{itemize}

\noindent and a fixed $\alphafe = +0.4$ and $\nfe=0$. Both the data and the models were continuum normalised with a running mean filter with a 30 pixel window. We limited the fit to the wavelength range $3700 - 5500$~\AA, where most of the features for extremely metal-poor stars are present. The spectra were shifted to rest-wavelength using the \ULySS radial velocities.

The resulting metallicity distribution is shown in red in the bottom panel of Figure~\ref{fig:ulyss}, without any additional quality cuts applied. The hard limit at $\feh_\FERRE = -2.0$ is due to the limit of the grid. The \ULySS and \FERRE distributions peak at roughly the same metallicities, but the \FERRE distribution has a larger tail towards lower metallicities -- as expected. 

We inspected the $>500$ fits in the resulting \FERRE-analysed sample with $\feh_\mathrm{\FERRE} < -3.0$ by eye, and identified a number of (previously unknown) stars of interest that could potentially have $\feh < -4.0$ or that looked very carbon-rich and extremely metal-poor ($\feh < -3.0$). Practically none of our candidates had parameters in the public DR6 catalogue. Most of our candidates had relatively low S/N, so follow-up spectroscopy was necessary to confirm their extremely or even ultra metal-poor nature. 

The full list of EMP candidates that we inspected is given in Table~\ref{tab:allcandidates}, with figures for all the spectral fits provided 
in the online supplementary materials.\footnote{The table and figures of FERRE fits for EMP candidates can temporarily be found at the following URL: \url{https://drive.google.com/drive/folders/1qLypkgyG6m-XbwPGpmNrBnXFQf70-0se?usp=sharing}, and they will be part of the online material with the published paper.} This candidate list with its derived parameters should not be used blindly since no quality cuts have been applied (on e.g. S/N, $\log(\chi^2)$ or parameter uncertainties), but it could be used in combination with the figures to select other EMP stars for follow-up. Stars may occur multiple times in this list if they have more than one LAMOST spectrum. 

\section{OSIRIS follow-up of EMP candidates}\label{sec:osiris}

We obtained GTC/OSIRIS observations for 11 of our most promising candidates ($16.9 < g < 17.9$) in Semester 2021A. We used OSIRIS in longslit mode with the 2500U grating, a 1 arcsec slit and 2x2 binning, resulting in spectra covering $3440 - 4610$~\AA at a resolving power $R \sim 2400$ (providing an instrument profile with a FWHM of $\sim 125$~\kms). We aimed for a S/N  of 40 at 4000~{\AA}, corresponding to exposure times of 3000s for stars of magnitude  $g\sim 17.5$. A summary of the observations is presented in Table~\ref{tab:observations}. Individual exposures of 1400, 1600 and 1800s were executed for different targets.  

\begin{table*}
\begin{center}
\caption{Summary of the OSIRIS observations: our reference for each star, the LAMOST and Gaia DR3 designations, positions, SDSS magnitudes in $u$ and $g$, total exposure time, signal to noise at two different wavelengths, number of observations and date that the spectra were observed. }
\resizebox{\textwidth}{!}{
 \begin{tabular}{|c c c c c c c c c c c|} 
 \hline
 Star \# & LAMOST  & Gaia DR3 source\_id & ra & dec & u\_mag & g\_mag & total exp  & S/N (@392/& Nobs & date observed \\ 
 & designation &   &   &   & [mag] & [mag] & time [s] & 450\,nm)  & & (DD-MM-2021) \\ 
 \hline\hline
LP1 & J002953.07+320229.9 & 	2862648994739368704	& 	00:29:53.07 	& 	32:02:29.9 	&  	18.49  	&  	17.58  	&  	2800	&  	33/79 	&	  	2	&	 	14-07 	\\		
LP2 & J131532.41+121107.4 & 	3736550805114696192	& 	13:15:32.41 	& 	12:11:07.4 	&  	18.33  	&  	17.44  	&  	2800	&  	57/127	 	& 	2	 	&	 11-04	 \\		
LP3 & J134510.95+424910.8 & 	1500794652785646976	& 	13:45:10.96 	& 	42:49:10.9 	&  	17.75  	&  	16.94  	&  	1400	&  	74/180	 	& 	1	 	&	 15-06	 \\		
LP4 & J142055.86+075308.7 & 	3673778479398720000	& 	14:20:55.87 	& 	07:53:08.7 	&  	17.96  	&  	17.04  	&  	1400	&  	62/100	 	& 	1	 	&	 15-06	 \\		
LP5 & J144714.22+163425.4 & 	1186662458446883328	& 	14:47:14.22 	& 	16:34:25.5 	&  	18.73  	&  	17.92  	&  	2800	&  	27/88 	&	  	2	&	 	10-06 	\\		
LP6 & J145214.98+160357.6 & 	1186397549159479424	& 	14:52:14.99 	& 	16:03:57.7 	&  	18.25  	&  	16.88  	&  	1400	&  	12/97 	&	  	1	&	 	10-06 	\\		
LP7 & J145611.30+161925.7 & 	1187873604865113472	& 	14:56:11.31 	& 	16:19:25.8 	&  	18.77  	&  	17.82  	&  	4600	&  	34/116	 	& 	3	 	&	 08-04	 \\		
LP8 & J161021.42+451247.5 & 	1386190837835580288	& 	16:10:21.43 	& 	45:12:47.6 	&  	18.50  	&  	17.63  	&  	5600	&  	30/79 	&	  	4	&	 	10-06/17-07 	\\	
LP9 & J162359.32+303740.8 & 	1318369490300720000	& 	16:23:59.32 	& 	30:37:40.9 	&  	18.46  	&  	17.52  	&  	5600	&  	24/78 	&	  	4	&	 	11-04/14-07 	\\	
LP10 & J212109.07+151328.7 & 	1783524305407324672	& 	21:21:09.06 	& 	15:13:29.0 	&  	18.15  	&  	17.15  	&  	4800	&  	38/145	 	& 	3	 &	 15-06/13-07/14-07 	\\
LP11 & J230209.39+302100.6 & 	1886140596052059392	& 	23:02:09.39 	& 	30:21:00.7 	&  	18.01  	&  	17.09  	&  	3600	&  	39/192	 	& 	2	 	&	 23-05/13-07	 \\	 [1.0ex]
 \hline
\end{tabular}}
\label{tab:observations}
\end{center}
\end{table*}

\begin{table*}
\begin{center}
\caption{Radial velocities, heliocentric distance, probability of the distance within 3$\sigma$ around the maximum of the distance PDF, maximum height from the plane, apocentric and pericentric distances, eccentricity, energy, the action vector are reported. [updated]} 
\resizebox{\textwidth}{!}{
\begin{tabular}{|c c c c c c c c c c c c c c c|} 
 \hline
Star \#  & RV & D & P$_D$ &$Z_\mathrm{max}$ & ${\rm R_{apo}}$ & ${\rm R_{peri}}$ & $\epsilon$ & E/10$^4$ & ${\rm J_{\phi}}$ & ${\rm J_{r}}$ & ${\rm J_{Z}}$  \\ 
  & [km s$^{-1}$] & [kpc] &  &[kpc] & [kpc] & [kpc] & & [kpc km$^{2}$ s$^{-2}$]&  [kpc km s$^{-1}$]&[kpc km s$^{-1}$]&[kpc km s$^{-1}$]  \\ \hline \hline
LP1 & $-18 \pm 15$ & $8.62 \pm 0.47$ & 0.14 & $15.6^{+4.9}_{-4.4}$ & $16.2^{+6.8}_{-3.1}$ & $2.4^{+1.3}_{-0.8}$ & $0.75^{+0.04}_{-0.03}$ & $-4.79^{+1.29}_{-0.86}$ & $-140^{+365}_{-417}$ & $829^{+177}_{-142}$ & $780^{+215}_{-329}$ \\ [0.7ex]
LP2 & $211 \pm 15$ & $6.88 \pm 0.15$ & 0.66 & $15.5^{+12.6}_{-4.7}$ & $24.3^{+31.4}_{-9.4}$ & $8.5^{+0.8}_{-0.7}$ & $0.52^{+0.19}_{-0.19}$ & $-2.76^{+2.17}_{-1.63}$ & $-1821^{+405}_{-434}$ & $17^{+1}_{-1}$ & $905^{+306}_{-153}$ \\ [0.7ex]
LP3 & $-162 \pm 15$ & $2.35 \pm 0.12$ & 0.99 & $8.0^{+1.6}_{-1.5}$ & $9.8^{+0.5}_{-0.4}$ & $2.4^{+0.7}_{-0.7}$ & $0.60^{+0.10}_{-0.10}$ & $-6.62^{+0.15}_{-0.13}$ & $368^{+197}_{-223}$ & $294^{+87}_{-76}$ & $501^{+188}_{-145}$ \\ [0.7ex]
LP4 & $-70 \pm 15$ & $6.03 \pm 0.10$ & 1.0 & $7.5^{+0.3}_{-1.3}$ & $7.8^{+0.2}_{-0.5}$ & $1.5^{+0.9}_{-0.6}$ & $0.66^{+0.12}_{-0.13}$ & $-7.58^{+0.36}_{-0.22}$ & $-15^{+207}_{-195}$ & $200^{+63}_{-60}$ & $556^{+159}_{-175}$ \\ [0.7ex]
LP5 & $-126 \pm 15$ & $3.25 \pm 0.16$ & 0.87 & $5.3^{+1.4}_{-0.8}$ & $10.1^{+0.4}_{-0.3}$ & $2.9^{+0.5}_{-0.5}$ & $0.54^{+0.06}_{-0.06}$ & $-6.50^{+0.14}_{-0.14}$ & $750^{+151}_{-151}$ & $269^{+56}_{-49}$ & $259^{+76}_{-64}$ \\ [0.7ex]
LP6 & $53 \pm 15$ & $22.7 \pm 1.4$ & 1.0 & $21.2^{+3.3}_{-3.6}$ & $21.7^{+5.7}_{-3.6}$ & $11.8^{+6.3}_{-3.5}$ & $0.30^{+0.06}_{-0.08}$ & $-3.02^{+1.16}_{-0.81}$ & $576^{+407}_{-203}$ & $198^{+43}_{-61}$ & $2480^{+770}_{-527}$ \\ [0.7ex]
LP7 & $-136 \pm 15$ & $8.05 \pm 0.18$ & 0.92 & $36.2^{+48.7}_{-22.3}$ & $52.8^{+135.8}_{-35.2}$ & $7.7^{+1.1}_{-0.7}$ & $0.74^{+0.16}_{-0.27}$ & $-0.85^{+4.15}_{-3.23}$ & $-1583^{+189}_{-99}$ & $38^{+1}_{-1}$ & $1382^{+889}_{-575}$ \\ [0.7ex]
LP8 & $-56 \pm 15$ & $3.62 \pm 0.23$ & 0.54 & $2.8^{+0.6}_{-0.5}$ & $9.2^{+0.4}_{-0.3}$ & $4.6^{+0.5}_{-0.5}$ & $0.33^{+0.05}_{-0.05}$ & $-6.48^{+0.18}_{-0.15}$ & $1215^{+107}_{-114}$ & $109^{+38}_{-27}$ & $112^{+25}_{-23}$ \\ [0.7ex]
LP9 & $-254 \pm 15$ & $7.25 \pm 0.14$ & 0.99 & $31.8^{+50.0}_{-19.8}$ & $39.9^{+90.7}_{-20.9}$ & $3.6^{+1.8}_{-0.7}$ & $0.83^{+0.08}_{-0.09}$ & $-1.61^{+3.56}_{-2.59}$ & $-726^{+196}_{-115}$ & $269^{+10}_{-10}$ & $903^{+943}_{-528}$ \\ [0.7ex]
LP10 & $-361 \pm 15$ & $5.79 \pm 0.10$ & 0.93 & $23.3^{+10.4}_{-11.3}$ & $52.4^{+115.2}_{-28.8}$ & $5.1^{+0.8}_{-0.4}$ & $0.83^{+0.10}_{-0.13}$ & $-0.70^{+3.49}_{-2.44}$ & $-1185^{+234}_{-182}$ & $191^{+10}_{-10}$ & $957^{+769}_{-500}$ \\ [0.7ex]
LP11 & $-68 \pm 15$ & $6.09 \pm 0.10$ & 0.99 & $12.5^{+6.6}_{-3.9}$ & $15.7^{+8.4}_{-3.9}$ & $10.5^{+0.7}_{-0.8}$ & $0.21^{+0.16}_{-0.09}$ & $-3.89^{+1.17}_{-0.78}$ & $1467^{+141}_{-116}$ & $58^{+255}_{-57}$ & $1084^{+465}_{-336}$
\\ \hline
\end{tabular}}
\label{tab:orbits}
\end{center}
\end{table*}

\subsection{Radial velocities}

\begin{figure*}
\begin{center}
{\includegraphics[width=0.85\hsize]{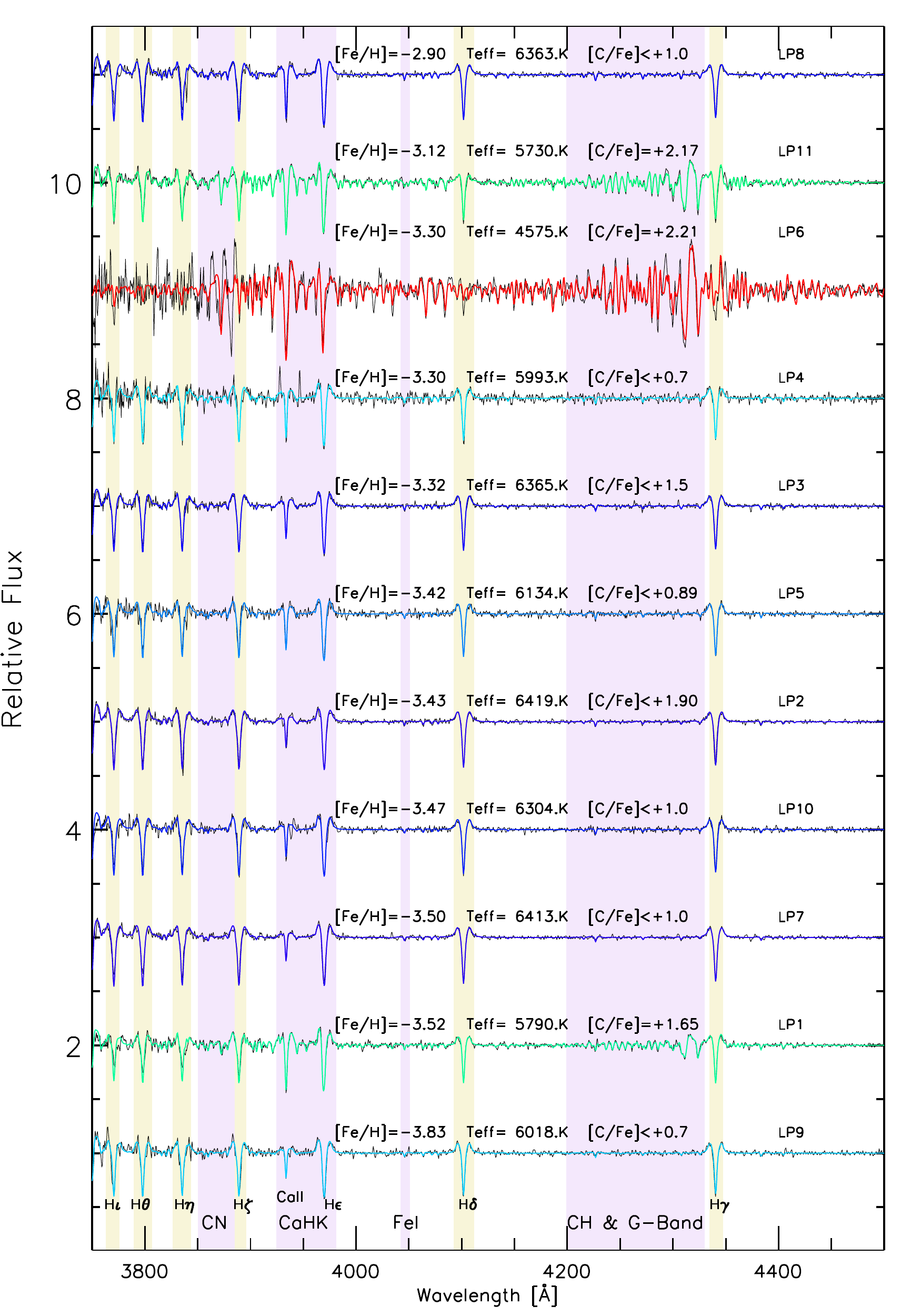}}
\end{center}
\caption{OSIRIS/GTC spectra (3750\,\AA-4500\,\AA) of our stellar sample
(black line) and the best fits calculated with \FERRE, colour-coded by \teff (the bluer the hotter) and sorted by decreasing \feh. The Balmer lines (yellow) and main metallic absorption features (purple) are high-lighted. Above each spectrum the metallicity, effective temperature and carbonicity are displayed. }
\label{fig:osiris}
\end{figure*}

Radial velocities (RVs) are derived using the cross-correlation technique. We have a high-quality GTC/OSIRIS spectrum of a bright extremely metal-poor star G64-12 ($\teff = 6463 \mathrm{K}, \logg = 4.26, \feh = -3.29, \cfe = +1.07$, \citealt{placco16} and references therein) from previous campaigns acquired with the same setup~\citep{agu17II,agu18}, that we use as a cross-correlation template. The OSIRIS spectra of both our targets and the template star are normalized with the same method, using a running mean filter with a width of 30 pixels.  
We built the cross-correlation function (CCF) with our own IDL-based automated code in the spectral range 3755-4455~{\AA} with a window of 3000~\kms. The main features in the template are the CaII H\&K lines, the HI lines of the Balmer series, and the G-band in carbon-enhanced stars (see Fig.~\ref{fig:osiris}). The normalization method produces a shape of the CCF profile that mimics the shape of all balmer lines in the warm template EMP star, which does not resemble a gaussian shape. We thus fit the CCF profile with a parabolic fit using the closest 6 points to the CCF peak. The statistical uncertainty of the centroid of the parabolic fit is typically below 1~\kms, significantly below the pixel size of $\sim 0.57$~{\AA}/pixel ($\sim 42$~\kms/pixel). The results of the OSIRIS spectra show intra-night RV variations with standard deviations below $\sim 7$~\kms, but RV variations from different nights with standard deviations in the range $3-20$~\kms. 

%%% \notea{I think these checks below were important for us to convince ourselves, but I don't think they need to be in the paper. }
% We found similar results (with differences below 5~\kms) using a different technique by normalizing the individual spectra with the task {\sc continuum} within IRAF environment using a cubic spline of third order and building the CCF with the IRAF task {\sc fxcor}. We did the same exercise using the \FERRE best-fit synthetic spectra (see Section~\ref{sec:ferre}) normalized with a running mean filter and found similar results with a mean difference of $-2.3$~\kms and a standard deviation of 7.8~\kms. 

We also derive the RV for the same stars from their LAMOST spectra (which typically have much lower S/N than the OSIRIS spectra), using the same technique to check the consistency with our OSIRIS RVs. The LAMOST spectrum of G64-12 is used as cross-correlation template and all spectra are normalized using a running mean filter with a width of 15 pixels of $\sim 1.38$~{\AA}/pixel ($\sim 81$~\kms/pixel).  
The CCF is built from the spectral range 3755-6755~{\AA}, which includes H$\alpha$ and H$\beta$, providing more stability to the CCF profile given the lower quality LAMOST spectra. We find a reasonable consistency when comparing to the OSIRIS results, with a mean difference of $-4.4$~\kms and a standard deviation of 15.9~\kms. 

For each target we adopt the weighted mean of the OSIRIS RVs derived from each individual spectrum and the corresponding error of the mean as the final RV. We apply an quadratically added uncertainty floor of 15~\kms to the RV uncertainties, which seems more realistic than the CCF uncertainties given the RV variations within and between different nights and the differences with the LAMOST RVs. This floor reflects the systematic RV uncertainty due to possible instrument flexures, pointing, guiding RV drifts, etc. 

\begin{table*}
\begin{center}
\caption{Table with adopted stellar parameters, with \teff, \feh and \cfe from FERRE and photometric \logg.}
\begin{tabular}{|l c c c c c c c c c |} 
 \hline
 Star \# & \teff & \logg & \feh & \cfe & corr.$^{2}$ &   \\
     & [K] & cgs &  & A(C)$_{\odot} =8.39^{1}$ & & \\ [0.5ex] 
 \hline\hline
LP1 & $5790 \pm 101$  & $4.83 \pm 0.22$   &  $-3.52 \pm 0.11$ &  $+1.65 \pm 0.21$  & \\
LP2 & $6419 \pm 108$  & $3.85 \pm 0.30$   &  $-3.43 \pm 0.12$ &  $<+1.90$  &   \\
LP3 & $6365 \pm 102$  & $4.57 \pm 0.08$   &  $-3.32 \pm 0.11$ &  $<+1.5 $  &  \\
LP4 & $5993 \pm 103$  & $3.65 \pm 0.28$   &  $-3.30 \pm 0.11$ & $<+0.70$  &   \\
LP5 & $6134 \pm 102$  & $4.59 \pm 0.37$   &  $-3.42 \pm 0.11$ &  $<+0.89 $  &   \\
LP6 & $4575 \pm 103$  & $1.00 \pm 0.20$   &  $-3.30 \pm 0.11$ &  $+2.21 \pm 0.21$  & $+0.24$  \\
LP7 & $6413 \pm 103$  & $3.85 \pm 0.39$   &  $-3.50 \pm 0.11$ & $<+1.0$  &   \\
LP8 & $6363 \pm 103$  & $4.47 \pm 0.19$   &  $-2.90 \pm 0.11$ &  $<+0.96 $  &   \\
LP9 & $6018 \pm 103$  & $3.70 \pm 0.21$   &  $-3.83 \pm 0.12$ & $<+0.70$  &  \\
LP10 & $6304 \pm 102$  & $3.75 \pm 0.14$   &  $-3.47 \pm 0.11$ &  $<+1.02 $  &   \\
LP11 & $5730 \pm 102$  & $3.53 \pm 0.24$   &  $-3.12 \pm 0.13$ &  $+2.17 \pm 0.22$  &  \\
 \hline
\end{tabular}

\vspace{0.05cm}
\label{tab:results}
\end{center}
$^{1}$Solar abundance adopted from \citet{asp05} \\
$^{2}$Evolutionary \cfe correction following \citet{placco14} \\
\end{table*}

\subsection{Distances}\label{sec:dist}
It has been widely demonstrated that simply inverting the parallax to infer the distance can lead to wrong results, and including additional priors and/or data improves distance estimates \citep[\eg][]{Bailer18, Bailer21, Anders22}. This is especially the case when the parallax has poor measurements, \ie $\varpi<0$ and/or $\varpi/\sigma_{\varpi}<20$. We therefore use a Bayesian approach to infer the distances for the stars in our sample. The probability distribution function (PDF), or posterior, is inferred following the method fully described in \citet{Sestito19}. Briefly, the likelihood is the product of the Gaussian distributions for the parallax and photometry. The prior takes into account a power-law stellar distribution \citep[see the halo prior in][]{Sestito19}, and, 
through a set of VMP ([M/H]$=-2.5$) MESA/MIST isochrones\footnote{\url{https://waps.cfa.harvard.edu/MIST/}} \citep{Dotter16,Choi16}, the knowledge that VMP stars are old ($11-13.8$ Gyr), low-mass ($<1\msun$), and distributed with a given IMF-based luminosity function in the CMD diagram. 
The zero-point offset has been applied to the {\it Gaia} EDR3 parallaxes \citep{Lindegren21} using the python \textsc{gaiadr3\_zeropoint}\footnote{\url{https://gitlab.com/icc-ub/public/gaiadr3\_zeropoint}} package. This method, widely used for chemo-dynamical investigations of VMP stars \citep[\eg][]{Sestito19,Sestito20,Venn20}, produces low uncertainties on the distances even in case of large parallax uncertainties. This is because the isochrones limit the possible distances for a star with a given colour to two different solutions, a dwarf and a giant solution, and nothing in between. The parallax would then typically prefer one of the two, or, in case of a very poor parallax measurement, the two peaks would be given a different probability. We calculate the probabilities following \citet{Sestito19}. For seven of the OSIRIS stars the probability of the main peak is larger than 92 per cent. For two stars it is 86 per cent (LP1, although for this star we adopt the less probable solution, see Section~\ref{sec:orb}) and 87 per cent (LP5), while for the remaining two stars it is 54 per cent (LP8) and 66 per cent (LP2).

\subsection{Orbital parameters}\label{sec:orb_desc}
The orbital parameters are inferred using  \textsc{Galpy}\footnote{\url{http://github.com/jobovy/galpy}}  \citep{Bovy15}. The code requires as input the inferred distances, the RVs, and the proper motions and coordinates from \textit{Gaia} (E)DR3. The total fixed gravitational potential that we adopt is the sum of a Navarro-Frenk-White dark matter halo \citep[][\textsc{NFWPotential}]{NavarroFrenkWhite97}, a Miyamoto-Nagai potential disc \citep[][\textsc{MiyamotoNagaiPotential}]{MiyamotoNagai} and an exponentially cut-off bulge (\textsc{PowerSphericalPotentialwCutoff}). All of the aforementioned potentials are usually invoked by the \textsc{MWPotential14} package. However, we adopt a more massive and up-to-date halo \citep{BlandHawthorn16}, with a mass of $1.2\times10^{12}\msun$ (vs. $0.8\times10^{12}\msun$ for  \textsc{MWPotential14}). 

For each star, we perform a Monte Carlo simulation with 1000 random draws on the input parameters to infer the orbital parameters and their uncertainties. In case of the proper motion components, we consider their correlation given the coefficients from  {\it Gaia} EDR3, drawing randomly with a multivariate Gaussian function.  The RVs (from the OSIRIS spectra) and coordinates are treated as a Gaussian. In order to account for possible systematics on the distances (e.g. due to the adopted isochrones and other assumptions), we assume a 15 per cent uncertainty on the distances. The integration time is set to 1 Gyr. The orbital parameters are inferred for both of the peaks in the distance PDFs.

The output orbital parameters are the Galactocentric Cartesian coordinates (X, Y, Z), the maximum distance from the Milky Way plane ${\rm Z_{max}}$, the apocentric and pericentric distances (${\rm R_{apo}}$, ${\rm R_{peri}}$), the eccentricity $\epsilon$, the energy E, and the spherical actions coordinates (${\rm J_{\phi}}$, ${\rm J_{r}}$, ${\rm J_{Z}}$).  Table~\ref{tab:orbits} reports the main orbital parameters from the most probable distance, except for star LP1 where we adopt the less probable distance (see Section~\ref{sec:orb}). The orbital parameters are discussed in Section~\ref{sec:orb}.

\subsection{Stellar parameters\label{sec:ferre}}

The OSIRIS data were analysed with \FERRE in a similar manner as the LAMOST spectra. For this analysis we use the more sophisticated Boender-Timmer-Rinnoy Kan \citep[BTRK,][]{boe82} global search algorithm and B\'ezier cubic interpolation. We use the same grid, except for the coolest star in the sample, for which we employ a similar grid that has been extended down to 4500~K (as used e.g. in \citealt{arentsen21}). Again we used a 30 pixel window for the running mean normalisation, suitable for OSIRIS resolution ($R = \lambda/\delta\lambda \sim 2400$). To avoid problems in the noisy blue region we only analyse the spectra in the range ($3750-4500$~\AA). 

\begin{figure}
\centering
\includegraphics[width=1.0\hsize,trim={0.0cm 1.0cm 0.0cm 0.0cm}]{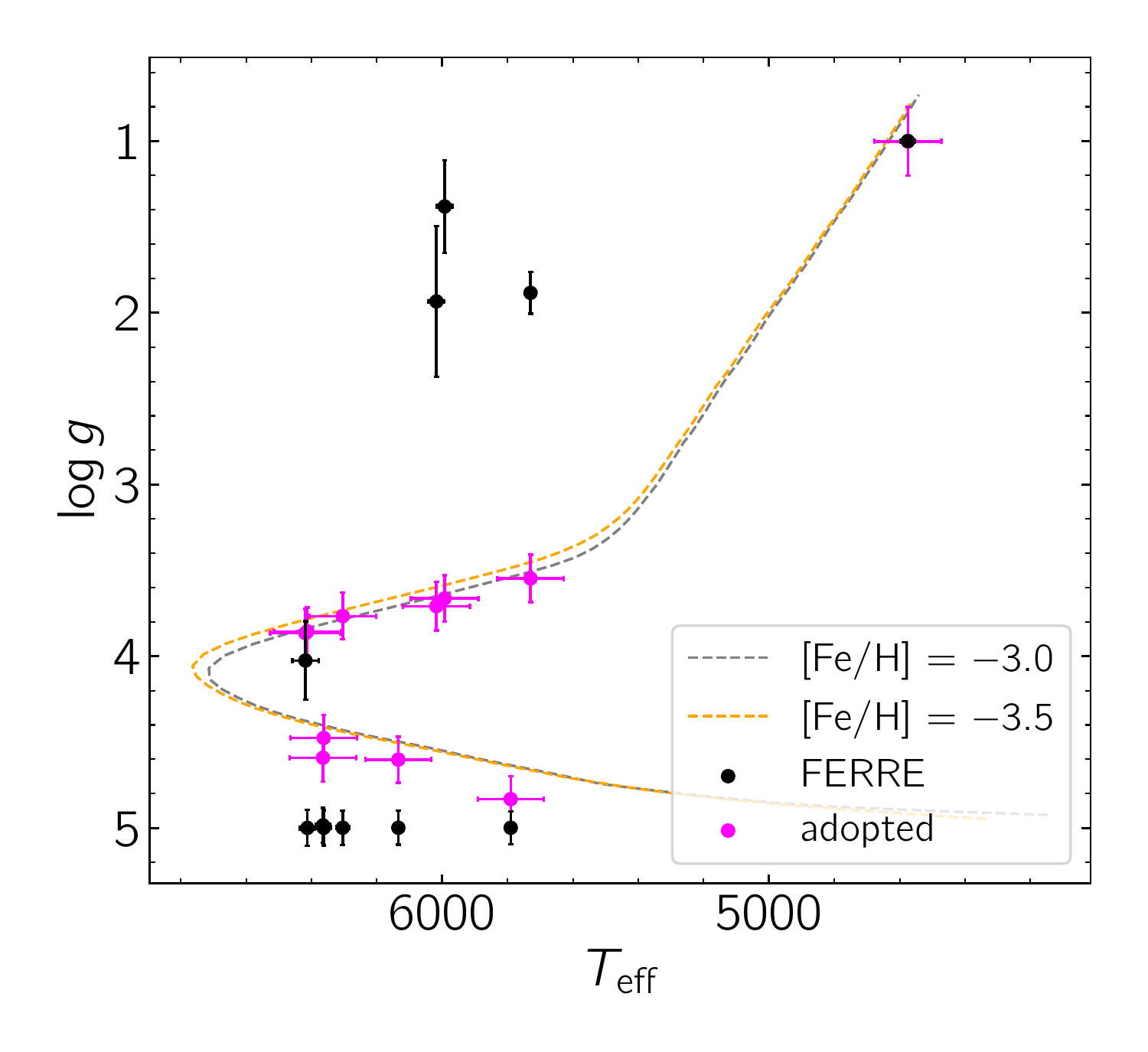}
\caption{Kiel diagram showing the pure FERRE stellar parameters (black) and the adopted stellar parameters and uncertainties (magenta). See the text for details. Also shown are Yonsei-Yale isochrones for two different metallicities (both with age = 12~Gyr, $\alphafe = +0.4$).}
\label{fig:tefflogg} 
\end{figure}

We found that for the warm stars in the sample (with $\teff > 5500$~K, which is all stars except for LP6), the \logg values that \FERRE finds are typically at the edges of the \FERRE grid, e.g. at $\logg = 5.0$ or $\logg < 2.0$, see the black points in Figure~\ref{fig:tefflogg}. This is likely the result of not much \logg information being present in these extremely metal-poor stars in the available wavelength range. Previous work on metal-poor stars with FERRE has shown that systematically offset \logg values strongly impact the derived \cfe \citep{aguado19,arentsen21}. Therefore we decided to adopt photometric \logg values for the warm stars, shown by the magenta points in Figure~\ref{fig:tefflogg}. These were inferred from the Stefan-Boltzmann equation, which needs as input the dereddened absolute G magnitude (derived using the \Gaia G-band magnitude, the 3D extinction map from \citealt{Green19} and the distances from Table~\ref{tab:orbits}), an estimate of the effective temperature, and the bolometric corrections on the flux \citep{Andrae18}. We adopt the FERRE effective temperature and its inflated uncertainty (see last paragraph of this subsection) in the calculation. We perform a Monte Carlo iteration with 1000 random draws on the input parameters. Each of them is described by a Gaussian distribution.

We run \FERRE again for the warm stars, fixing the \teff to the previously derived \FERRE value and \logg to the photometric values values, while letting \feh and \cfe free. The final spectral fits are shown in Figure~\ref{fig:osiris} and a summary of the results is provided in Table~\ref{tab:results}. The differences between the original \FERRE run and the run with fixed \teff and \logg are small for the metallicities, with the adopted \feh being higher by 0.07~dex with a standard deviation of 0.06~dex. The differences for \cfe are also small for the stars with original $\logg > 4$ and measured \cfe (see next section), they are 0.05 on average, with a standard deviation of 0.09~dex. However, for the one star with measured carbon and \FERRE $\logg < 3$ (LP11), the new \cfe is 0.7~dex lower.

There are three stars (LP4, LP7 and LP9) that have very high \FERRE internal \feh uncertainties of $0.5-1.0$~dex when calculated by inverting the covariance matrix (our original approach). This could be attributed to some negative/zero fluxes in blue end of the OSIRIS data. To avoid this issue we recalculated the internal \FERRE uncertainties using a Monte Carlo simulation. We performed 50 experiments and use the dispersion on the derived [Fe/H] and [C/Fe] as the uncertainty following \citet{agu17II}. As a result of that the issue with the large uncertainties was fixed for the three problematic stars, and the uncertainties for the other stars remain the same within 0.01$-$0.02\,dex. We adopt the Monte Carlo internal uncertainties.

To provide the final uncertainties for the stellar parameters, we add estimates of the external uncertainties from a previous analysis of EMP stars with \FERRE \citep{agu17II} to our internal \FERRE uncertainties. These are 100~K, 0.1 and 0.2~dex for \teff, \feh and \cfe, respectively. For \feh and \cfe we adopt the internal uncertainties from the first \FERRE run, because the second run does not properly reflect the real uncertainties since it fits only two of the four parameters. For \logg, we adopted the uncertainties from the photometric determination for the warm stars, and for the coolest star we quadratically added 0.2~dex of external uncertainties \citep{agu17II} to the internal \FERRE uncertainty. The results are shown in Table~\ref{tab:results}.
 
\subsection{Carbon determination}
 
Deriving carbon abundance from low-resolution data of EMP stars is non-trivial. Our employed grid is suitable for the analysis of CEMP stars, since carbon-enhancement was not only considered in the spectral synthesis step but also in the ATLAS stellar models \citep{sbordone07}. This is crucial because high carbon abundances can significantly impact the stellar atmospheres. The grid of models has been used successfully to derive carbon abundances in several works \citep[e.g.][]{agu17I, agu17II, aguado19, arentsen21, arentsen22}, although there are some differences with other synthetic grids that can lead to systematic differences in derived carbon abundances \citep{arentsen22}. This is likely related to the use of different codes, line lists and assumptions (e.g. different \nfe abundances). 

The ability of the FERRE code to detect -- and successfully fit -- carbon absorption features from low-resolution data strongly depends on a)\,$\teff$ (and $\logg$ to a lesser extent), b)\,the carbon abundance, and c)\,the SNR of the spectra. In our sample there are three stars (LP1, LP6, and LP11) that fulfil the sensitivity criteria derived by \citet{aguado19} based on these parameters, all of them have $\teff<6000$~K and show strong CH absorption features. For these objects we derived $\rm [C/Fe]=+1.65/+2.21/+2.17$ respectively, with reasonable uncertainties ($\sim0.2$\,dex). For the other stars we can only provide upper limits on the carbon abundances. The carbon results are summarised in Table~\ref{tab:results}.

The object with the lowest \teff in our sample, LP6, shows clear CN features at $\sim3885$\,\AA that our best fit is not able to reproduce, although the CH \& G-band fit is good (see Fig. \ref{fig:osiris}, red spectrum). The reason for this is that our FERRE synthetic spectral library assumes $\nfe=0.0$ for all stellar models. 
Querying the high-resolution spectroscopy compilation in the JINAbase \citep{abohalima18} for stars with $-3.5 < \feh < -3.0$, we find that all of those with measured nitrogen abundances have $\nfe > 0$, and stars with $\cfe > +2.0$ typically have $1.5 < \nfe < 3.0$. This is very different from the assumed \nfe in the \FERRE grid, and can explain why the CN band for LP6 is much stronger in the data than in the model fit. However, the fit reproduces quite well the \ion{Ca}{ii} at 3933\,\AA and several other \ion{Fe}{i}, \ion{Ti}{ii}, and \ion{Sr}{ii} lines in the 4040$-$4080\,\AA region. Additionally, the majority of the carbon information is significantly concentrated around the G$-$band (4200-4330\,\AA) and our fit is good in that area. Therefore, we conclude that the CN absorption features in the blue are not significantly affecting the best fit for this object.

The carbon abundance of evolved giants decreases with decreasing \logg due to mixing processes, especially in metal-poor stars \citep{gratton00,placco14}. We estimate the evolutionary carbon correction for the most evolved star in our sample (LP6, the only star that should be affected by this effect) using the web calculator\footnote{\url{https://vplacco.pythonanywhere.com/}} by V.M. Placco, and find it to be +0.24~dex.

\begin{figure*}
\centering
\includegraphics[width=0.5\hsize]{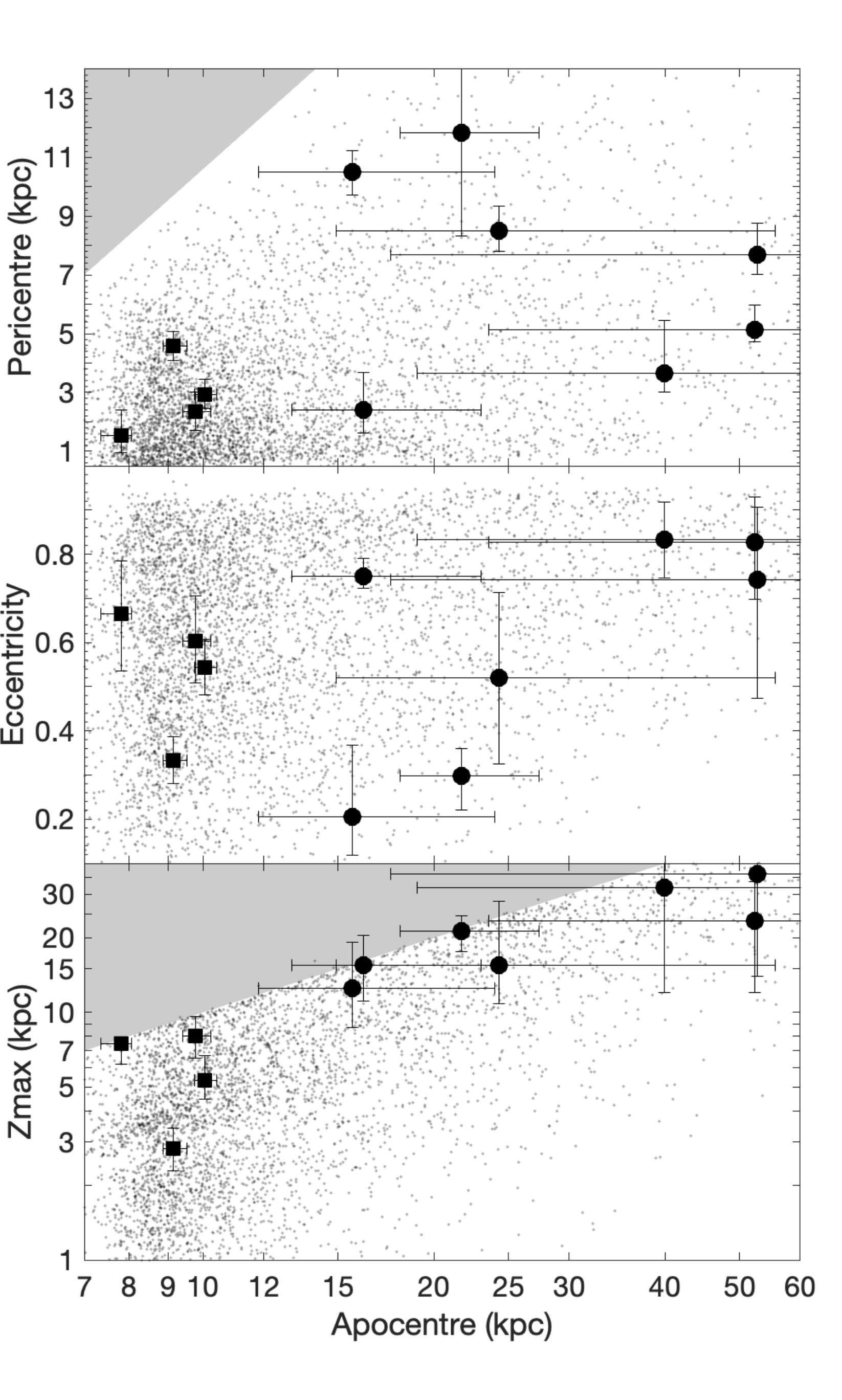}
\includegraphics[width=0.44\hsize, trim={ 0.cm .4cm .cm 0.7cm},clip]{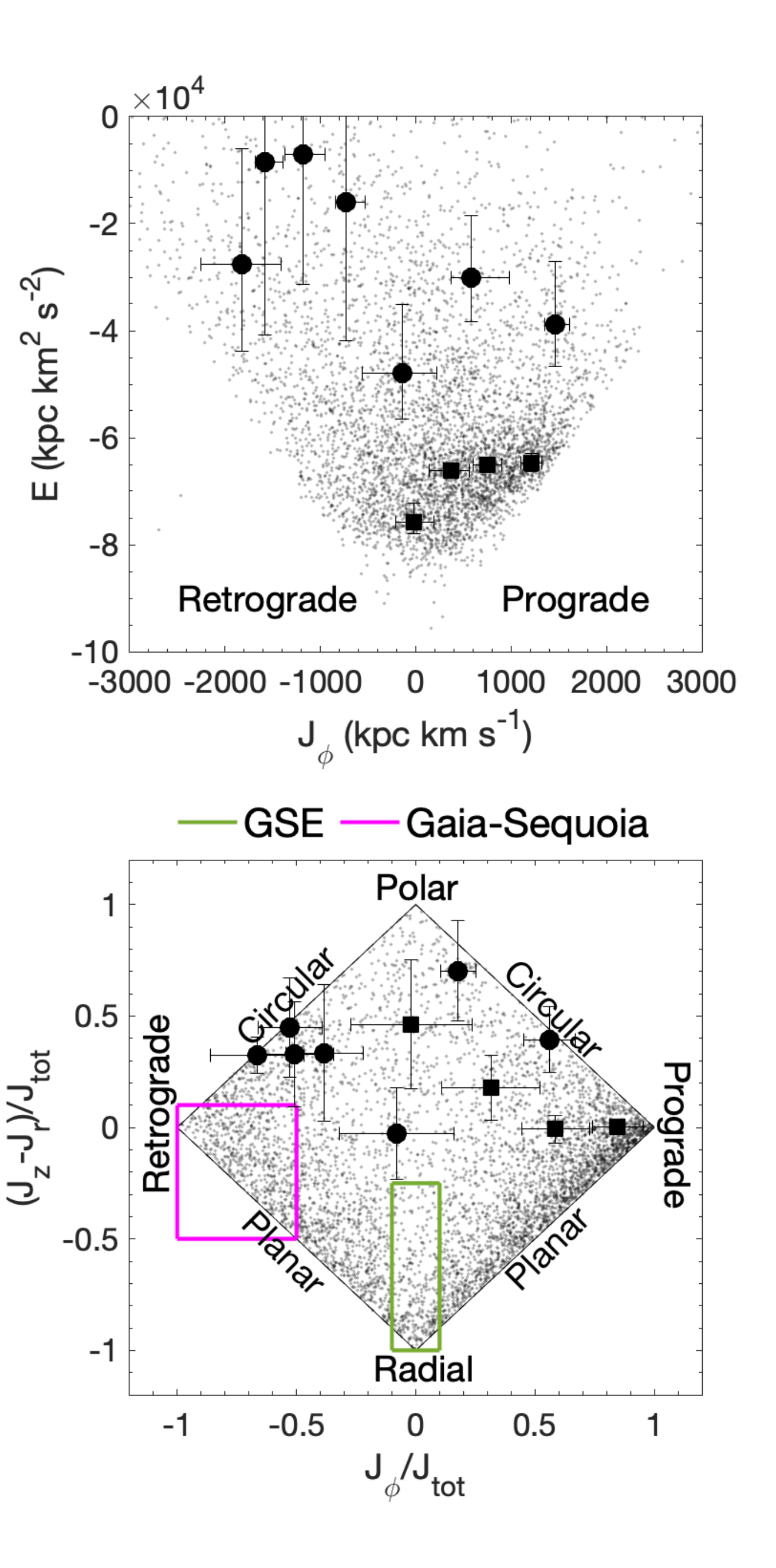}
\caption{Orbital parameters.  Three left panels:  pericenter, eccentricity, and  maximum distance from the Milky Way plane as a function of the apocentric distance. The grey-shaded areas denote the forbidden region in which the Z$_{\rm{max}} > {\rm R_{apo}}$ or ${\rm R_{peri}} > {\rm R_{apo}}$. Upper right panel: Energy vs. rotational component of the action, J$_{\phi}$. Bottom right panel: Action space; the y-axis is the difference between the vertical and radial component of the action, while the x-axis is the rotational component; axes are normalised by J$_{\rm{tot}} = |\rm{J}_{\phi}| + \rm{J}_{\rm{r}} +\rm{J}_{\rm{Z}} $. The inner  (${\rm R_{apo}} <11$ kpc) and the outer (${\rm R_{apo}} >15$ kpc)  groups are squares and circles, respectively. Green and magenta solid lines in the bottom right panel denotes the regions of Gaia-Sausage/Enceladus \citep[][]{Belokurov18,Helmi18} and Sequoia \citep{Barba19, myeong2019}, respectively. Grey small dots in the background of all panels are VMP stars studied in \citet{Sestito20}, in which the orbital parameters have been inferred with the same potential as this work.}
\label{fig:apopar} 
\end{figure*}

\section{OSIRIS sample results}\label{sec:results}

The derived properties for our 11 OSIRIS stars are summarised in Tables~\ref{tab:orbits} and \ref{tab:results}. In this section, we will use these parameters to study the Galactic orbital properties of our sample, to study the carbon-enhanced metal-poor stars in our sample, and to make a comparison with a recent LAMOST catalogue that includes VMP stars.

\subsection{Orbital properties}\label{sec:orb}

Here we discuss the orbital parameters for our EMP OSIRIS sample. We adopted the results for the most probable distance solution (see Section~\ref{sec:dist}), except for LP1 for which the most probable solution leads to an unbound orbit -- we therefore prefer the less probable distance solution for this star. The five panels in Figure~\ref{fig:apopar} display the main orbital parameters typically used to classify the kinematic properties of stars. The three panels on the left-hand side show the pericentric distance, the eccentricity, and the maximum height from the plane as a function of the apocentric distance. The right-hand two panels display the energy vs. the rotational component of the action (top) and the action space (bottom). The sample appears to split into two broad populations in the Z$_{\rm{max}}$ vs. apocenter and the E vs J$_{\phi}$ panels -- one that inhabits the inner region of the Milky Way (${\rm R_{apo}} \lesssim 10$ kpc) and one that reaches the outer part Milky Way halo (${\rm R_{apo}} \gtrsim 15$ kpc). We mark these with black squares and  circles, respectively. 

The first group is composed of four stars with apocentric distances of $\sim 7-10$ kpc. Three of them (LP3, LP4, LP5) have pericentres that bring them into the spatial region of the Milky Way bulge (${\rm R_{peri}}<3$ kpc). The remaining one, LP8, has a higher pericenter (${\rm R_{peri}}\sim4.5$ kpc) and is among the lowest eccentricity stars in the sample ($\epsilon \sim 0.3$) -- its ${\rm Z_{max}}<3.0$ kpc and positive angular momentum indicate the star is moving in a prograde orbit relatively close to the plane of the Milky Way. All stars in this group are prograde, with the exception of LP4, which has a very high eccentric orbit ($\epsilon\sim0.7$), and almost no rotation ($J_{\phi}/J_{\rm{tot}}\sim 0$). These extremely metal-poor inner halo stars may be connected to very first Milky Way halo building blocks, the ancient Galactic disk and/or the chaotic (but slightly rotating) pre-disk Milky Way. 

The second group is composed of the remaining seven stars with orbits compatible with outer halo stars. Three of them, LP1, LP9 and LP10, have pericentric distances in the range $2.0<{\rm R_{peri}}<5.5$ kpc, the other four, LP2, LP6, LP7 and LP11, have larger pericentric distances. From the action space of Figure~\ref{fig:apopar}, it is evident that none of our targets is clearly kinematically associated with GSE (green box) or Sequoia (magenta box). One of the stars, LP1 (sitting near the centre of the action diamond), could still have belonged to the GSE progenitor since it has high eccentricity ($\sim0.75$) and is not far out of the GSE box. Previous works have associated some stars in this region with GSE \citep[e.g.][]{yuan20} or shown that in simulations there are GSE stars on a variety of orbits larger than the typical selection boxes \citep[e.g.][]{naidu21,amarante22}. A possible association of LP11 (the most prograde star in the outer halo group) can be made with the Helmi stream \citep{helmi99}, as it is sits in a similar region of the action diamond and the E-$J_{\phi}$ space \citep[see e.g.][]{yuan20} and has strong vertical motion ($J_z = 1084$~kpc\,km\,s$^{-1}$), consistent with the very polar orbit of the Helmi stream. Association with other halo-substructures (such as the dynamically tagged groups of VMP stars by \citealt{yuan20} and others) is difficult due to the relatively large uncertainties on the orbital parameters for most stars. The majority of our stars were likely brought into the Milky Way in smaller accretion events. 

High-resolution spectroscopic observations would be needed to determine the detailed chemo-dynamical properties of the stars in this work. They would provide better RVs to derive more precise orbital parameters and more importantly detailed chemical abundances, from different nucleosynthetic production channels, which are needed to better characterise the formation sites and origins of the stars in our sample.

\begin{figure}
\centering
\includegraphics[width=1.0\hsize,trim={0.0cm 1.0cm 0.0cm 0.0cm}]{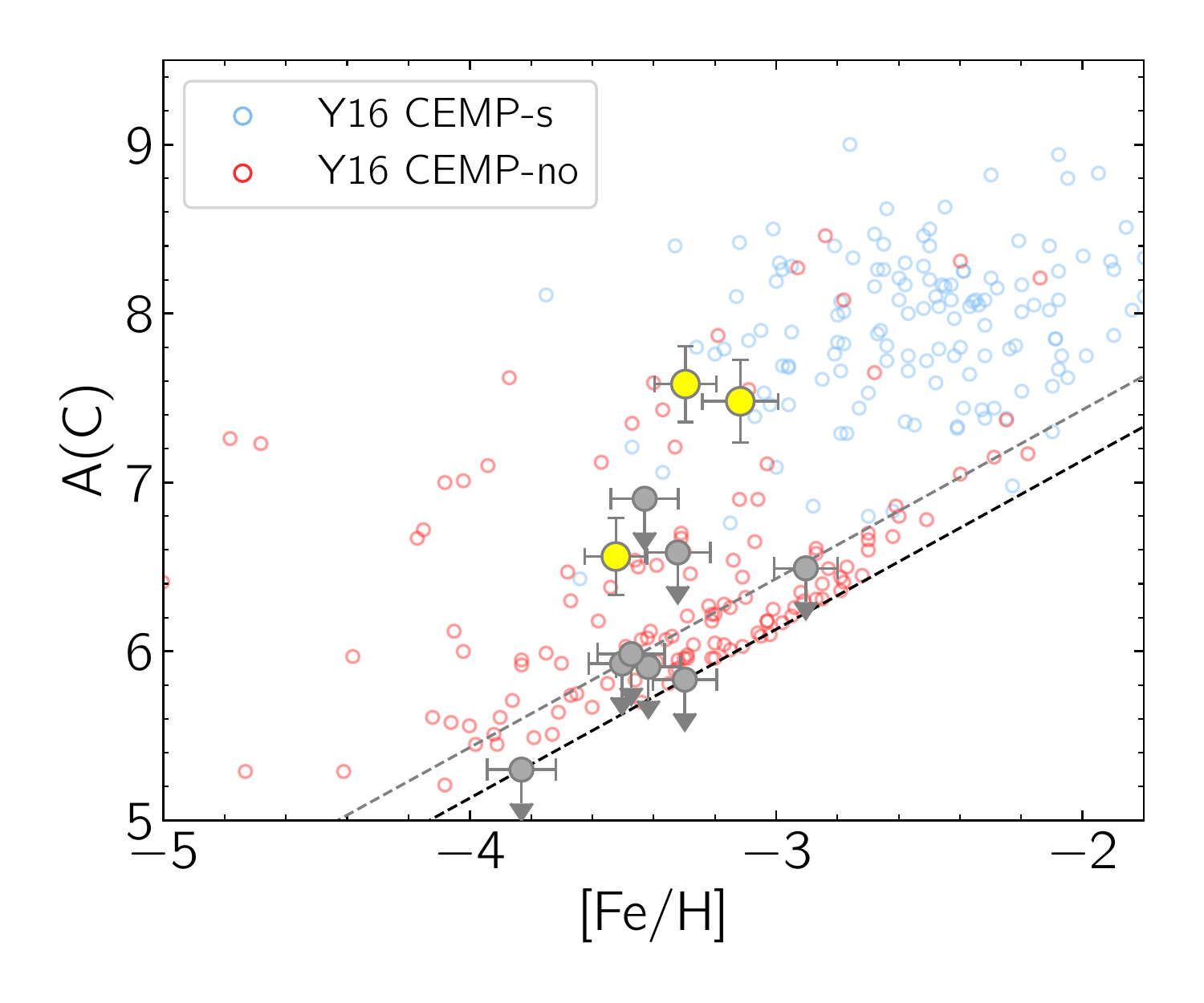}
\caption{\feh versus \ac (corrected for evolutionary effects) for the stars in our sample (large yellow symbols, and grey symbols for upper limits) and the CEMP stars in the \citet{yoon16} compilation (small symbols colour-coded by CEMP type). The uncertainties on \ac are the quadratic sum of the adopted uncertainties on \feh and \cfe. The black and grey dashed lines indicate the limits of $\cfe = +0.7$ and $+1.0$, respectively.
}
\label{fig:AC} 
\end{figure}

\subsection{CEMP stars}\label{sec:CEMP}

Following the \citet{aoki07} definition of CEMP stars ($\cfe > +0.7$), three of our stars can be classified as carbon-enhanced: LP1, LP6 and LP11. For two other objects (LP4 and LP9, with $\teff\sim6000$~K but no clear features within the G band), we were able to provide an informative upper limit of $\cfe<+0.7$, making these carbon-normal stars. The other six targets (LP2, LP3, LP5, LP7, LP8, and LP10) are relatively warm ($\teff>6100$~K) and the absence of CH absorption features only allow us to provide upper limits that are larger than $\cfe = +0.7$, according to the sensitivity criteria from \citet{aguado19}. We do not derive the fraction of CEMP stars in our sample, since the preselection was strongly biased.

Since we do not have estimates of any s-process element abundances for our sample\footnote{There are two relatively strong lines of Sr and Ba in our wavelength coverage, but the combination of resolution, S/N and extremely low metallicities of the stars do not permit their detection.}, we cannot constrain the types of CEMP stars in our sample using that method. However, CEMP-s and CEMP-no stars also have different distributions in their metallicities and carbon abundances \citep[e.g.][]{spite13, bonifacio15, yoon16}. We can use this to make a preliminary classification of CEMP stars. Figure~\ref{fig:AC} presents the $\feh-\ac$\footnote{$A$(C) $= \log{\epsilon (C)} = \log(N_C/N_H) + 12$, with A(C)$_{\odot} =8.39$ from \citet{asp05}} diagram of the stars in our sample,  together with a compilation of CEMP stars from \citet{yoon16}. The two most carbon-rich CEMP stars in our sample (LP6 and LP11) are on the border between the CEMP-no and CEMP-s regions. The third (LP1) lies in the CEMP-no region of the diagram, as well as the other stars with \cfe upper limits.

All three CEMP stars have large apocentres ($> 20$~kpc), and the two most carbon-rich CEMP stars also have the highest pericentres in our sample ($>8$~kpc). As discussed above, these are indications that they likely came into the Milky Way in a relatively small dwarf galaxy. 
Previous work has suggested that the fraction of CEMP-no compared to CEMP-s stars is larger in the outer halo than in the inner halo \citep{yoon18, lee19}, as well as in smaller halo building blocks \citep{yoon19, zepeda22}.
This is additional indirect evidence that the two most carbon-rich stars in our sample are more likely to be CEMP-no.

If LP6 and LP11 are CEMP-s stars, they are among the lowest metallicity CEMP-s stars known. If they are CEMP-no stars, they are among the highest-\ac CEMP-no stars known. There are not that many literature stars in this region, so it would be interesting to do further higher resolution follow-up of these two stars to investigate their nature.

\subsection{LAMOST DR8 VaC comparison}

A new analysis of the LAMOST DR8 spectra was published in a value-added-catalogue (VaC) by \citet{wang22}, employing neural networks to derive stellar parameters (\teff, \logg and \feh). They train one of the neural networks on stars of all metallicities in the PASTEL catalogue \citep{soubiran10}, and another network only on metal-poor stars ($\feh < -1.5$) to improve their \feh estimates for VMP stars. They claim that the metallicities in their VMP catalogue are reliable down to $\feh \sim -3.5$. 

\begin{figure}
\centering
\includegraphics[width=1.0\hsize,trim={0.0cm 1.0cm 0.0cm 0.0cm}]{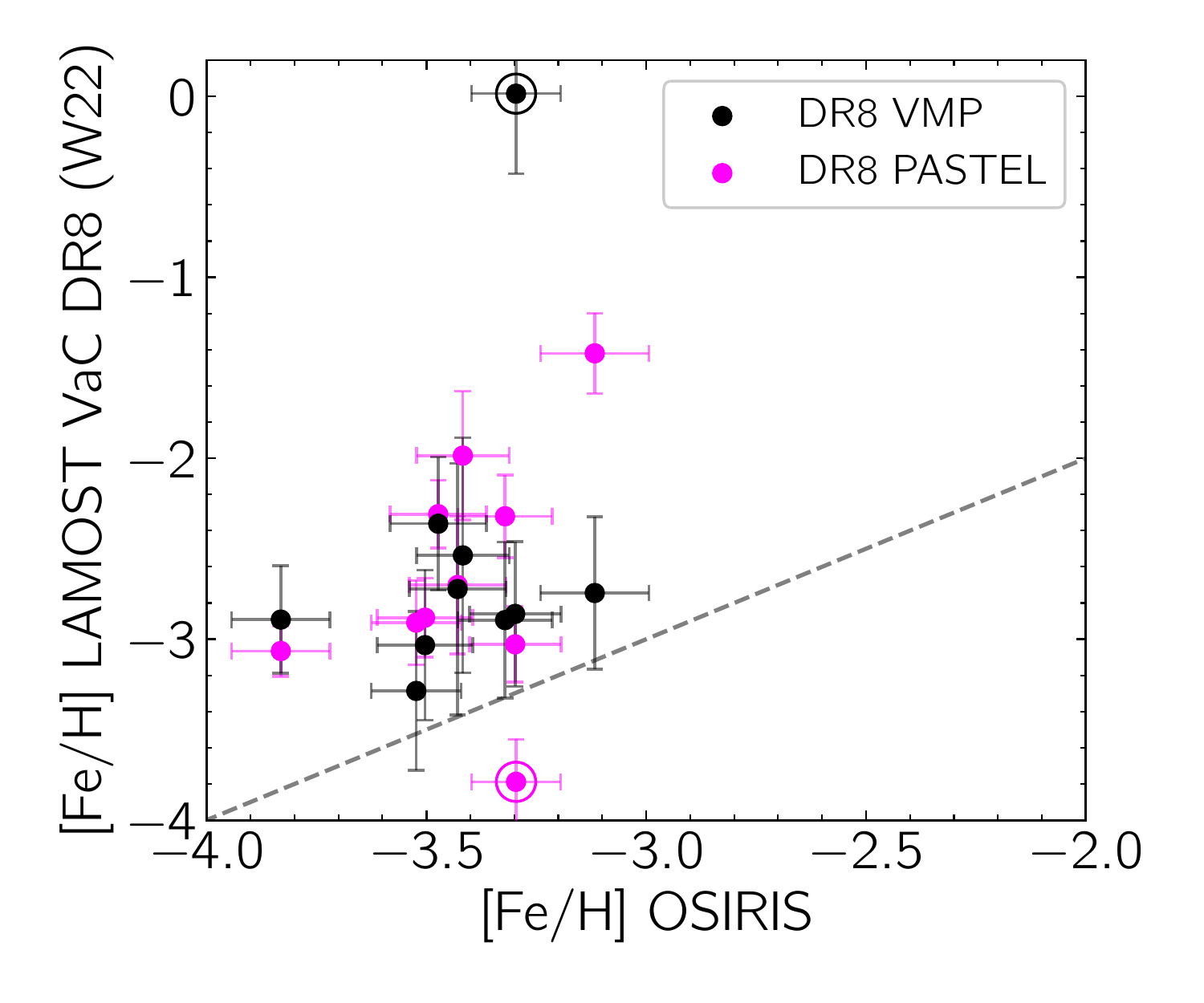}
\caption{Comparison between our derived metallicities from the OSIRIS spectra and those from the \citet{wang22} LAMOST DR8 value-added-catalogue. The points are colour-coded by the version of the neural network applied to the DR8 data, and the cool CEMP star in our sample (LP6) has been highlighted with a large circle. The bisector is indicated with a grey-dashed line.}
\label{fig:fehcomp} 
\end{figure}

Ten out of our eleven OSIRIS stars have stellar parameters in the DR8 VaC (the only star absent is our most metal-rich star, LP8, with $\feh_\FERRE = -2.9$). We present the comparison between the DR8 VaC metallicities and the metallicities derived in this work in Figure~\ref{fig:fehcomp}. The very carbon-enhanced cool star LP6 has extreme metallicities in both the PASTEL and VMP catalogues, which is not unexpected since the spectrum is dominated by carbon features and this is not taken into account in the \citet{wang22} analysis. Focusing on the $\feh_\mathrm{VMP}$ estimates, the other stars are all found to have systematically higher metallicities compared to our analysis, mostly between $-3.0 < \feh_\mathrm{W22} < -2.3$. Since we are using spectra of much higher SNR and we are employing a dedicated analysis method for extremely metal-poor (and/or carbon-enhanced) stars, we conclude that some caution should be taken with the \citet{wang22} VMP catalogues for $\feh_\mathrm{W22} < -2.5$. We further note that more EMP stars may be hidden in large catalogues, especially among stars with low S/N spectra.

\section{Summary} \label{sec:conclusions}

In this work, we employed the combination of metallicity-sensitive photometry from the \Pristine survey \citep{starkenburg17} and the large low-resolution spectroscopic LAMOST database to identify promising ultra metal-poor and/or carbon-enhanced extremely metal-poor candidates. We analysed $\sim 7500$ LAMOST spectra for targets with $\feh_\Pristine < -2.5$ and $g < 18$, finding success rates of stars with $\feh_\mathrm{spec} < -2.5$ between $34\% - 50\%$, depending on the applied quality cuts. We inspected all the fits with $\feh_\mathrm{spec} < -3.0$ to identify candidates for follow-up, and we release this full list together with figures of the best fits (see Section~\ref{sec:fullferre}).

We observed eleven of the most exciting candidates (mostly with low LAMOST S/N) using OSIRIS at the GTC. We analysed the higher S/N medium-resolution OSIRIS spectra ($R \sim 2400$) using the \FERRE code to derive \teff, \feh and \cfe, adopting \logg from photometry. The metallicities for the eleven stars range from $\feh = -2.9 \pm 0.1 $ to $-3.8 \pm 0.2$, with a mean $\feh = -3.4$. We set out to identify UMP stars, but none of the targets had $\feh < -4.0$ -- such stars are indeed incredibly rare. Our selection of (carbon-enhanced) extremely metal-poor stars, however, was still very efficient. 

For three out of the eleven stars we were able to derive carbon abundances, for the others we derived upper limits -- two of which are constraining and classify the stars as carbon-normal. Given their \feh, \ac \,\,and orbital properties, all three CEMP stars are likely part of the CEMP-no category, although the two most carbon-rich objects lie in an underpopulated region, where there are both CEMP-no and CEMP-s stars in the literature. Further follow-up is necessary to understand the physical processes causing the carbon-enhancement in these stars. 

We derive orbital properties using the OSIRIS radial velocities, \Gaia proper motions and distances based on photometry and parallaxes from \Gaia combined with MIST isochrones, integrating orbits in the \texttt{MW-Potential14} with a more massive halo. We find that four of the stars have inner halo kinematics, with three of them on prograde orbits. The other seven stars have orbits more consistent with the outer halo. None of the stars in our sample are confidently associated with previously known substructures/accretion events, partly due to uncertainties on the orbital parameters. 

Ongoing and upcoming spectroscopic surveys are so large that it is crucial to have general automatic analyses of the spectra, but doing this well for extremely metal-poor stars is a challenge. They are only a small subset, hence pipelines are often not optimised for them, and their spectra are challenging to analyse due to weak spectral features and/or peculiar chemical abundances. It will remain important to do dedicated metal-poor analyses in the future. Adding additional information like metallicity-sensitive photometry as in this work could uncover hidden promising candidates at the lowest metallicities.

\section*{Acknowledgements}
We thank the reviewer for their valuable comments, which helped improve the paper. The authors thank Carlos Allende Prieto, Carmela Lardo and Lyudmila Mashonkina, as well as the rest of the \Pristine collaboration, for their support of this paper and/or their useful comments. 

AA, NFM and ZY gratefully acknowledge support from the European Research Council (ERC) under the European Unions Horizon 2020 research and innovation programme (grant agreement No. 834148). 
AA acknowledges support from the Herchel Smith Fellowship at the University of Cambridge and the Fitzwilliam College Isaac Newton Trust Research Fellowship. 
DA acknowledges support from the European Research Council (ERC) Starting Grant NEFERTITI H2020/808240.
FS thanks the Dr. Margaret "Marmie" Perkins Hess postdoctoral fellowship for funding his work at the University of Victoria.
JIGH acknowledges financial support from the Spanish Ministry of Science and Innovation (MICINN) project PID2020-117493GB-I00 and also from the Spanish MICINN under 2013 Ram\'on y Cajal program RYC-2013-14875. 
NFM and ZY acknowledge support from the French National Research Agency (ANR) funded project “Pristine” (ANR-18-CE31-0017), NFM also acknowledges funding from CNRS/INSU through the Programme National Galaxies et Cosmologie and through the CNRS grant PICS07708. 
ES acknowledges funding through VIDI grant ``Pushing Galactic Archaeology to its limits'' (with project number VI.Vidi.193.093) which is funded by the Dutch Research Council (NWO).
PJ acknowlegdes support from the Swiss National Foundation. 

Based on observations made with the Gran Telescopio Canarias (GTC), installed at the Spanish Observatorio del Roque de los Muchachos of the Instituto de Astrofísica de Canarias, on the island of La Palma.

Based on observations obtained with
MegaPrime/MegaCam, a joint project of CFHT and
CEA/DAPNIA, at the Canada-France-Hawaii Telescope
(CFHT) which is operated by the National Research Council
(NRC) of Canada, the Institut National des Sciences de
l’Univers of the Centre National de la Recherche Scientifique
of France, and the University of Hawaii.

Guoshoujing Telescope (the Large Sky Area Multi-Object Fiber Spectroscopic Telescope LAMOST) is a National Major Scientific Project built by the Chinese Academy of Sciences. Funding for the project has been provided by the National Development and Reform Commission. LAMOST is operated and managed by the National Astronomical Observatories, Chinese Academy of Sciences.

Funding for the Sloan Digital Sky 
Survey IV has been provided by the 
Alfred P. Sloan Foundation, the U.S. 
Department of Energy Office of 
Science, and the Participating 
Institutions. SDSS-IV acknowledges support and 
resources from the Center for High 
Performance Computing  at the 
University of Utah. The SDSS 
website is www.sdss.org. SDSS-IV is managed by the 
Astrophysical Research Consortium 
for the Participating Institutions 
of the SDSS Collaboration including 
the Brazilian Participation Group, 
the Carnegie Institution for Science, 
Carnegie Mellon University, Center for 
Astrophysics | Harvard \& 
Smithsonian, the Chilean Participation 
Group, the French Participation Group, 
Instituto de Astrof\'isica de 
Canarias, The Johns Hopkins 
University, Kavli Institute for the 
Physics and Mathematics of the 
Universe (IPMU) / University of 
Tokyo, the Korean Participation Group, 
Lawrence Berkeley National Laboratory, 
Leibniz Institut f\"ur Astrophysik 
Potsdam (AIP),  Max-Planck-Institut 
f\"ur Astronomie (MPIA Heidelberg), 
Max-Planck-Institut f\"ur 
Astrophysik (MPA Garching), 
Max-Planck-Institut f\"ur 
Extraterrestrische Physik (MPE), 
National Astronomical Observatories of 
China, New Mexico State University, 
New York University, University of 
Notre Dame, Observat\'ario 
Nacional / MCTI, The Ohio State 
University, Pennsylvania State 
University, Shanghai 
Astronomical Observatory, United 
Kingdom Participation Group, 
Universidad Nacional Aut\'onoma 
de M\'exico, University of Arizona, 
University of Colorado Boulder, 
University of Oxford, University of 
Portsmouth, University of Utah, 
University of Virginia, University 
of Washington, University of 
Wisconsin, Vanderbilt University, 
and Yale University.

This work has made use of data from the European Space Agency (ESA) mission
{\it Gaia} (\url{https://www.cosmos.esa.int/gaia}), processed by the {\it Gaia} Data Processing and Analysis Consortium (DPAC, \url{https://www.cosmos.esa.int/web/gaia/dpac/consortium}). Funding for the DPAC has been provided by national institutions, in particular the institutions participating in the {\it Gaia} Multilateral Agreement.

%%%%%%%%%%%%%%%%%%%%%%%%%%%%%%%%%%%%%%%%%%%%%%%%%%
\section*{Data Availability}

The LAMOST spectra used in this work are public. Our EMP candidate list is available in Table~\ref{tab:allcandidates}, and all relevant data for the OSIRIS stars is available in Tables~\ref{tab:observations}~$-$~\ref{tab:results}. These tables will also be available at the CDS. The OSIRIS spectra will be shared on reasonable request to the authors. 



\bibliographystyle{mnras}
\bibliography{pristine-osiris} % 

%%%%%%%%%%%%%%%%%%%%%%%%%%%%%%%%%%%%%%%%%%%%%%%%%%

%%%%%%%%%%%%%%%%% APPENDICES %%%%%%%%%%%%%%%%%%%%%

% \appendix

% \section{Some extra material}

% If you want to present additional material which would interrupt the flow of the main paper,
% it can be placed in an Appendix which appears after the list of references.

%%%%%%%%%%%%%%%%%%%%%%%%%%%%%%%%%%%%%%%%%%%%%%%%%%

% Don't change these lines
\bsp	% typesetting comment
\label{lastpage}
\end{document}